\documentclass[PRM,twocolumn,groupedaddress,nofootinbib,superscriptaddress]{revtex4-1}
\usepackage{graphicx}
\usepackage{dcolumn}
\usepackage{bm}
\usepackage{times}
\usepackage{epstopdf}
\usepackage[version=3]{mhchem} 
\usepackage{xcolor}

\usepackage{mathrsfs}
\usepackage{lipsum}
\usepackage{amsmath}
\usepackage{amsfonts}
\usepackage{array}
\usepackage{color}
\usepackage{ulem}

\newcommand{\degreesC}{\,$^{\circ}$C}
\newcommand{\Tc}{$T_{\rm c}$}

\begin{document}

\title{Electronic and topological properties of the van der Waals layered superconductor PtTe}

\author{Michael A. McGuire}
\email{McGuireMA@ornl.gov \\  \\ Notice: This manuscript has been authored by UT-Battelle, LLC under Contract No. DE-AC05-00OR22725 with the U.S. Department of Energy. The United States Government retains and the publisher, by accepting the article for publication, acknowledges that the United States Government retains a non-exclusive, paid-up, irrevocable, world-wide license to publish or reproduce the published form of this manuscript, or allow others to do so, for United States Government purposes. The Department of Energy will provide public access to these results of federally sponsored research in accordance with the DOE Public Access Plan (http://energy.gov/downloads/doe-public-access-plan). }
\author{Yun-Yi Pai}
\author{Matthew Brahlek}
\author{Satoshi Okamoto}
\author{R. G. Moore}

\affiliation{Materials Science and Technology Division, Oak Ridge National Laboratory, Oak Ridge, Tennessee 37831 USA}

\begin{abstract}
We report the crystal growth and structural and electronic properties of superconducting, van der Waals layered PtTe. Easily cleavable crystals with a plate-like morphology consistent with the layered structure were grown from a platinum rich flux. A consistent determination of $T_c = 0.57$\,K is made from the onset of diamagnetism, the zero of resistivity, and the midpoint of the heat capacity jump. The observed behavior is consistent with type-II superconductivity, with upper critical field at $T=0$ estimated using the Werthamer-Helfand-Hohenberg theory to be 143 and 65\,Oe for fields out of and in the plane, respectively. The heat capacity discontinuity is close to the weak coupling BCS value. Density functional theory calculations and analysis of the electronic structure finds that PtTe is a topological semimetal with numerous surface states, but suggests the superconducting state itself may be topologically trivial. Angle resolved photoemission spectroscopy reveals a normal-state Fermi surface in remarkable agreement with theory, and confirms the overall topological nature of the material by experimental identification of the surface bands. Together, these findings identify PtTe as an interesting example of a cleavable, topological, and superconducting material.
\end{abstract}

\maketitle

\section{Introduction}

Pseudo two dimensional materials comprising layers that are weakly bound together by van der Waals forces play important roles in many active areas of condensed matter physics. This includes magnetism, topology, optoelectronics, and spintronics \cite{huang2020emergent, li2019intrinsic, sierra2021van}. Monohalides with the ZrCl structure type have been rediscovered as interesting examples of such materials \cite{zhou2019discovery, song2021van, li2021theoretical}. These compounds have a double layer of metal cations in a highly reduced formal oxidation state. The double layer is bonded to halide anions above and below, producing slabs that stack together by van der Waals bonding to form a rhombohedral crystal structure. Materials with this structure have recently attracted attention as electrides \cite{zhou2019discovery}, where electrons confined between the metal layers act as anions, and as 2D magnets when the cations are rare-earth metals \cite{song2021van, li2021theoretical}. It is also noteworthy that the rare-earth containing ZrCl-type materials may be stabilized by interstitial hydrogen \cite{meyer1986synthetic}. The van der Waals layered structures of these compounds and their metallic nature make this an interesting class of materials.

PtTe is the only non-halide reported to adopt the ZrCl structure type \cite{cenzual1990overlooked}. The strong spin orbit coupling and pseudo-2D nature identify it as a potentially interesting topological material. Indeed, the Topological Materials Database \cite{bradlyn2017topological, vergniory2019complete, vergniory2021all} identifies several non-zero topological indices for this compound. Almost 60 years ago, in Ref. \citenum{Raub1965}, the single phrase ``PtTe is supercondcuting at 0.59\,K'' is printed about this compound. The compound was noted to be orthorhombic there, but likely it refers to ZrCl-type PtTe and the full symmetry had simply not been identified at the time. Little if any additional information appears to have been reported in the meantime, although thin films have recently been grown \cite{zhang2021growth}. Thus, PtTe provides potential to study combined topological and superconducting behaviors in a cleavable material. Interesting electronic properties have been reported in some chemically related phases (note that direct structural analogs of PtTe are limited to the halides mentioned above). PdTe, adopting the NiAs structure type, is a superconductor below 4.5\,K \cite{karki2012pdte}, and \ce{CdI2}-type palladium compound \ce{PdTe2} is superconducting with $T_c$ near 1.7\,K and hosts Type-II Dirac Fermions \cite{noh2017experimental, leng2017type} while \ce{CdI2}-type \ce{PtTe2} is also a Dirac semimetal \cite{PtTe2} and is converted to a relatively high temperature ferromagnet when doped heavily with chromium \cite{CrPtTe2}.

Here we report a thorough experimental and theoretical study of the electronic properties of superconducting PtTe. Crystals grown from a platinum rich flux have a plate-like morphology consistent with the layered structure. We find a consistent determination of $T_c = 0.57$\,K from the onset of diamagnetism, the zero of resistivity, and the midpoint of the heat capacity jump. The observed behavior is consistent with type-II superconductivity, with relatively low critical fields. Density functional theory calculations of the electronic structure and analysis of the associated topological properties finds that PtTe is a topological semimetal with numerous surface bands, but the superconducting state is expected to be topologically trivial. Angle resolved photoemission spectroscopy reveals a normal-state Fermi surface in remarkable agreement with the band structure calculations, and confirm the topological nature of the material by experimental identification of surface bands.

\section{Experimental details and methods}

\subsection{Crystal growth}

PtTe is the most Pt rich binary compound on the Pt-Te phase diagram. It is reported to be a line compound that decomposes at a peritectic reaction near 935\degreesC\ into a liquid and \ce{Pt3Te4}. The eutectic between Pt and PtTe is at about 870\degreesC\, which presents a suitable temperature window for a PtTe crystal growth via flux method using a Pt rich melt. Crystals studied here were grown from a melt of composition \ce{Pt60Te40} contained using alumina Canfield crucible sets \cite{Canfield2016}, comprising a ``growth'' crucible into which the reactants are loaded, an alumina frit designed to separate crystals from melt, and a ``catch'' crucible inverted on top of the frit. The loaded crucibles were sealed inside evacuated silica ampoules, heated to heated to 1000\degreesC\ and held for 24 hours before cooling to 920\degreesC\ at 1-2 \degreesC/hr. At this temperature the ampoules were inverted into a centrifuge to remove the molten flux from the crystals. We note that cooling the melt to 900\degreesC\ resulted in freezing of the flux in the growth crucible before it could be decanted, and that using a melt of composition \ce{Pt54Te46} produced a mixture of crystals containing primarily PtTe with some \ce{Pt3Te4}. When using a \ce{Pt60Te40} mixture and decanting at 920\degreesC\ as described above, the growth crucibles were found to contain many PtTe crystals with platelike morphologies, with lateral dimensions up to 8 mm and typical thicknesses of 0.1$-$0.3 mm (Fig. \ref{fig:structure}c). The crystals can be mechanically cleaved into thinner plates relatively easily and are exfoliable using tape. X-ray diffraction from the as grown and cleaved facets gives reflections consistent with the layer spacing along the rhombohedral c axis (hexagonal setting) in the reported crystal structure. Semiquantitative energy dispersive x-ray spectroscopy analysis indicated a Pt/Te ratio of 0.96(3), consistent with the composition PtTe within expected accuracy. Selected PtTe crystals were annealed at 400\degreesC\ for several days to see how this might affect the superconducting transition temperature. No change in \Tc\ was detected in resistivity measurements.

\subsection{Structure and physical property measurements}

Powder x-ray diffraction was performed using a PANalytical X'Pert Pro MPD diffractometer with monochromatic Cu-K$_{\alpha1}$ radiation, and analyzed using Fullprof \cite{Fullprof}. Single crystal x-ray diffraction was measured using Mo K$_{\alpha1}$ with a Bruker D8 Quest diffractometer with the sample held at 200\,K in a nitrogen cold stream. Semiempirical absorption corrections were applied using SADABS and structural refinements were performed using ShelX \cite{ShelX}. Commercial cryostats from Quantum Design (PPMS, Dyancool, MPMS-XL, MPMS-3) were used for heat capacity, electrical transport, and magnetization measurements, with their associated He-3 options for measurements below 1.8\,K. Lower temperature resistivity measurements were carried out in an Oxford Triton dilution refrigerator with Stanford Research SR860 lock-in amplifiers using a small current (50 or 100\,$\mu$A) to minimize heating.

\subsection{Angle resolved photoemission spectroscopy}

Angle resolved photoemission spectroscopy (ARPES) measurements were performed on cleaved bulk crystal samples.  Measurements were performed in a lab based system using a Scienta DA30L hemispherical analyzer with a base pressure of $P = 5 \times 10^{-11}$ Torr and a base temperature of $T \sim 7$ K.   Samples were mounted to sample holder using silver paste with a ceramic post glued to the top of the crystal.  The samples were cleaved in vacuum and at base temperature by knocking off the post.  Samples were illuminated with linearly horizontal polarized light using an Oxide $h\nu = 11$ eV laser system with nominal $60$ $\mu$m spot diameter~\cite{He2016}.  The intensity of the laser allowed for an analyzer configuration with pass energy of 2 eV and 0.2 mm slit resulting in a total energy resolution $\sim 2$ meV and momentum resolution $\sim  0.007$ \AA$^{-1}$ while still yielding sufficient electron counts for efficient measurement times.  The analyzer deflection mode was used to map out sections of the electronic structure within a $\pm 12^\circ$ deflection angle while partial maps centered at steeper angles were combined to construct the entire Fermi surface in the first Brillouin zone (BZ).

\subsection{First principles electronic structure calculations}

Density functional theory (DFT) calculations used the projector augmented wave method \cite{Kresse1999} with the generalized gradient approximation in the parametrization of Perdew, Burke, and Enzerhof \cite{Perdew1996} for exchange correlation as implemented in the Vienna {\it ab} initio simulation package (VASP) \cite{Kresse1996}. For both Pt and Te standard potentials were used (Pt and Te in the VASP distribution). In most cases, we used a $12 \times 12 \times 12$ {\bf k}-point grid and an energy cutoff of 500 eV. The spin-orbit coupling was included, to enable better comparison with ARPES data, but the $+U$ correction was not included because PtTe is an itinerant nonmagnetic system.

\section{Results and discussion}

\subsection{Crystal structure}
\begin{figure}
\begin{center}
\includegraphics[width=3.0in]{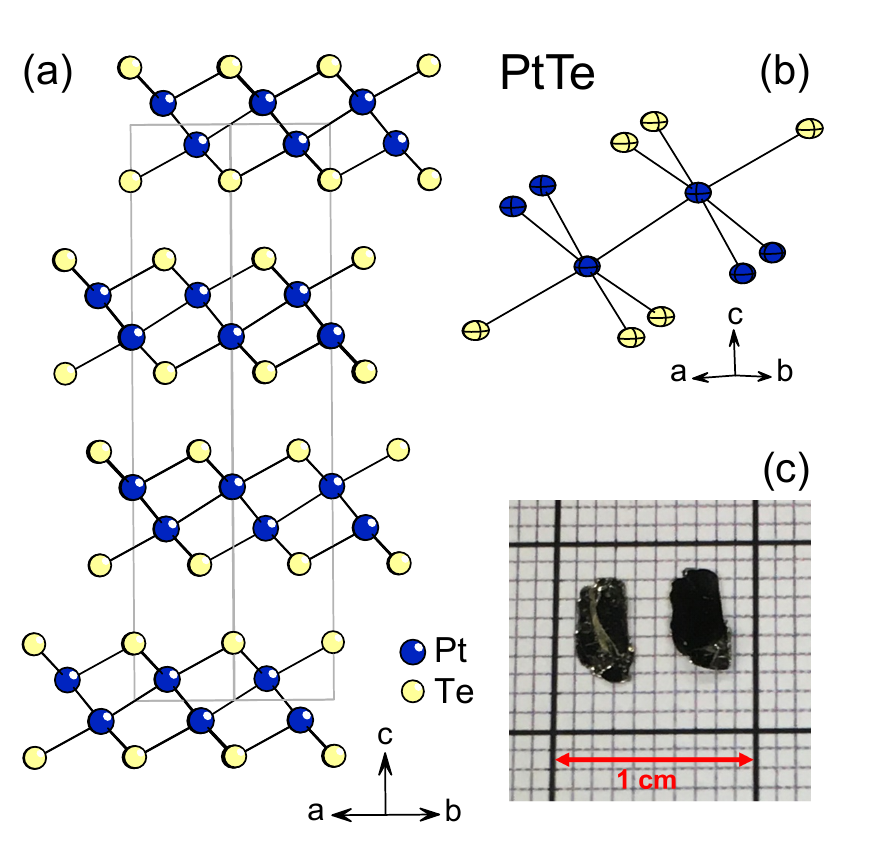}
\caption{\label{fig:structure}
The crystal structure of PtTe. (a) A view down the [110] direction in the hexagonal setting  emphasizing the layered nature of the crystal structure. (b) A view showing the octahedral coordination of the Pt atoms (blue) by three other Pt atoms and three Te atoms (yellow) with refined atomic displacement parameters (95\%) reflected in oblate shape of the atoms. (c) A typical crystal obtained from flux growth, cleaved in two perpendicular to the c axis, showing an as grown surface on the left and the cleaved surface on the right.
}
\end{center}
\end{figure}

Our single crystal and powder x-ray diffraction measurements confirm the rhombohedral structure of PtTe reported in Ref. \citenum{cenzual1990overlooked}. The structure contains four-atom-thick slabs separated by a relatively large gap (Fig. \ref{fig:structure}a), suggesting van der Waals interactions between the slabs and consistent with the platelike morphology and cleavability of the crystals (Fig. \ref{fig:structure}c). Each slab contains a double layer of Pt capped above and below by Te layers. All the layers are close packed in the plane. The Pt atoms are in octahedral coordination, bonded to three Te atoms and three Pt atoms (Fig. \ref{fig:structure}b). Lattice parameters, atomic coordinates, and equivalent isotropic displacement parameters are listed in Table \ref{tab:xrd}. Including vacancies or mixed occupancy on either of the atomic positions in the structure did not improve the refinement results, indicating that the compound is close to stoichiometric, consistent with semiquantitative energy dispersive spectroscopy measurement results. Powder x-ray diffraction confirmed the large crystals were single phase PtTe; however, significant strain and texture was induced by grinding this layered material. An attempt to minimize this was made by annealing the ground powder for three days at 400\degreesC. Results of Rietveld refinement of the data collected at room temperature are compared with the single crystal diffraction results (200\,K) in Table \ref{tab:xrd}, and show reasonable agreement. For comparison, the lattice parameters reported in Ref. \citenum{cenzual1990overlooked} are a = 3.963\,{\AA} and c = 19.98\,{\AA}.  Further details of our diffraction measurements and results are included in the Supplemental Material \cite{supp}.

As noted above, PtTe crystals cleave easily in the plane of the structural layers. Mechanical cleavage/exfoliation was accomplished with a blade as done for the crystal in Fig. \ref{fig:structure}c, and samples are also easily cleaved using tape. This suggests monolayer or few-layer specimens of PtTe may be obtainable. The shortest interlayer distance between Te atoms spanning the van der Waals gap in PtTe is 3.54\,{\AA}. The most familiar two dimensional tellurium compounds that have been exfoliated to down to a few or one layer tend to have somewhat larger Te-Te distances across their van der Waals gaps. The 2D valleytronic and superconducting material \ce{MoTe2} \cite{yang2015robust, rhodes2021enhanced} and the 2D topological insulator \ce{WTe2} \cite{fei2017edge}, have interlayer Te-Te distances of 3.9\,{\AA} \cite{brown1966crystal}. In ferromagnetic \ce{CrGeTe3} \cite{gong2017discovery}, the interlayer distance is 4.0\,{\AA} \cite{carteaux1995crystallographic}. Finally, the ferromagnets \ce{Fe3GeTe2} and \ce{Fe5GeTe2} \cite{Fei2018, Ohta2020} have shortest interlayer Te-Te distances of 3.7\,{\AA} and 3.8\,{\AA}, respectively \cite{deiseroth2006fe3gete2, stahl2018van, may2019ferromagnetism}. While the interlayer Te-Te distance in PtTe is shorter than these examples, it is comparable to the interchain distance in the 1-D van der Waals structure of elemental Te, which has been mechanically exfoliated to produce few-nanometer-thick wires \cite{churchill2017toward}. In the trigonal structure of Te, the atoms within a chain are separated by 2.83\,{\AA} and the closest van der Waals contact between chains is 3.49\,{\AA} \cite{adenis1989reinvestigation}. Few layer specimens of \ce{Bi2Te3}, which has an interlayer distance of 3.65\,{\AA} \cite{feutelais1993study}, have also been reported  \cite{teweldebrhan2010atomically, shahil2012micro}. It is interesting to compare these distances with tellurium's reported van der Waals radius of 2.1\,{\AA} \cite{bondi1964van, mantina2009consistent}, corresponding to a Te$-$Te van der Waals distance of 4.2\,{\AA}. The ditellurides of Mo and W and \ce{CrGeTe3} come somewhat close to this value.

\begin{table}
\caption{\label{tab:xrd} Crystal structure refinement results for PtTe (space group $R\overline{3}m$, hexagonal setting) using single crystal x-ray diffraction data collected at 200\,K and powder x-ray diffraction data collected at room temperature.}
\setlength{\tabcolsep}{5mm}
\begin{tabular}{lcc}					
\hline
	&	single crystal	&	powder	\\
\textit{a} ({\AA})	&	3.9612(4)	&	3.9575(6)	\\
\textit{c} ({\AA})	&	19.815(2)	&	19.819(3)	\\
Pt-z	&	0.3695(1)	&	0.3685(2)	\\
Te-z	&	0.0986(1)	&	0.0986(3)	\\
Pt-U(eq) ({\AA}$^2$)	&	0.007(1)	&	--	\\
Te-U(eq) ({\AA}$^2$)	&	0.007(1)	&	--	\\
R-values	&	R1 = 0.071	&	Rp = 12.5	\\
	&	wR2 = 0.167	&	RwP = 16.1	\\
\hline				
\end{tabular}	
\end{table}

\subsection{Normal state properties}

\begin{figure}
\begin{center}
\includegraphics[width=3.5in]{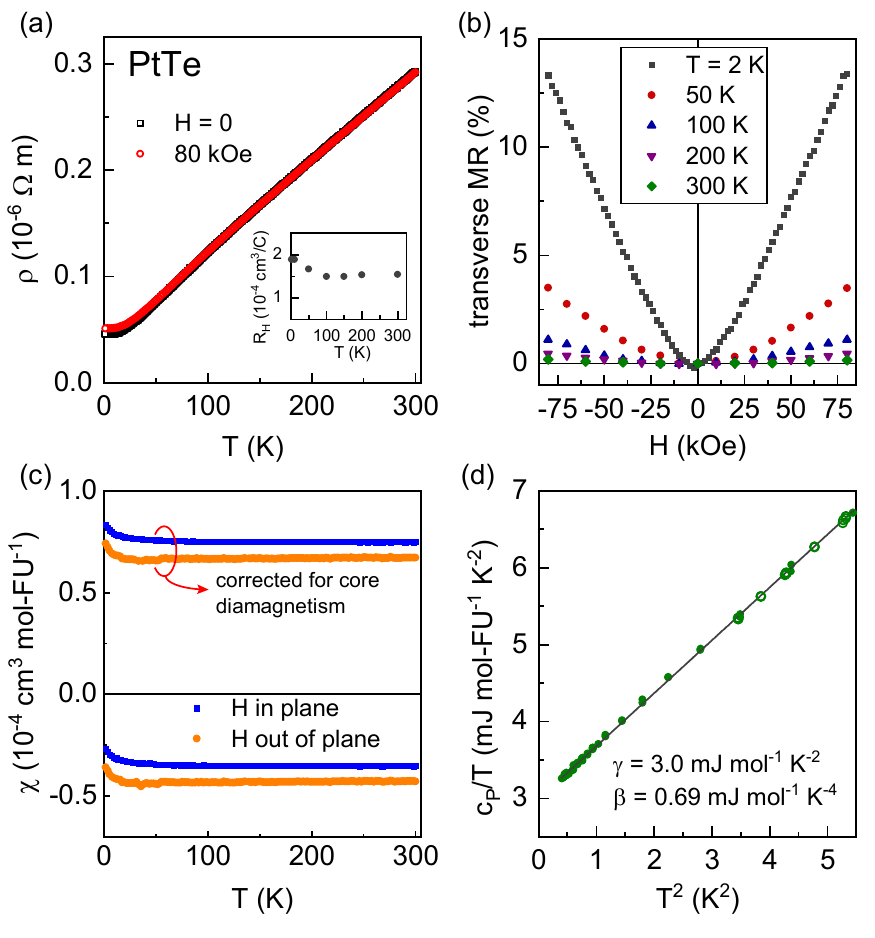}
\caption{\label{fig:normal}
Normal state properties of PtTe. (a) Electrical resistivity measured in zero magnetic field and in 80\,kOe, with Hall coefficient shown in the inset. (b) Transverse magnetoresistance measured with the applied field out of the plane (along the c axis) and the current in the plane (perpendicular to the c axis). (c) Magnetic susceptibility measured with a field of 10\,kOe in the plane and out of the plane. The measured data is shown along with data that has been corrected for core diamagnetic contribution of -40$\times 10^{-6}$\,cm$^3$/mol and -70$\times 10^{-6}$\,cm$^3$/mol for Pt$^{2+}$ and Te$^{2-}$, respectively \cite{corediamagnetism}. (d) Low temperature heat capacity showing a $\gamma T + \beta T^3$ temperature dependence, with open and closed symbols representing data from two separate measurements.
}
\end{center}
\end{figure}

Results of electrical transport and magnetization measurements of PtTe crystals between 2 and 300\,K are summarized in Fig. \ref{fig:normal}. The temperature dependence of the electrical resistivity (Fig. \ref{fig:normal}a) shows typical metallic behavior. Upon cooling, a crossover from a linear temperature dependence at high temperature to a stronger power law is seen near 40\,K (about 30\% of the Debye temperature as determined by heat capacity, see below), consistent with electron-phonon scattering. Between about 8 and 40\,K the data can be described by $\rho(T) = \rho_0 + \rm{A} \textit{T}^n$, with a fitted value of 1.87(3) for the exponent $n$. The resistivity has a very weak temperature dependence between 2 and 8\,K. The residual resistivity ratio $\rho(300 K)/\rho(2 K)$ in zero field is 6.4. The Hall coefficient is shown in the inset of Fig. \ref{fig:normal}. It is positive over the entire temperature range studied here, indicating transport dominated by holes. Interpretation of the Hall effect data can be complicated by the presence of many bands at the Fermi level in this material. However, the electronic structure calculations presented below indicate the bands crossing the Fermi level are hole bands, in agreement with the Hall effect data. The measured Hall voltage was linear with applied field from -80 to 80\,kOe at all of the measurement temperatures.

%

The transverse magnetoresistance (MR) measured with the field along the c-axis (normal to the cleavage plane) and the current in the ab plane is shown in Fig. \ref{fig:normal}b. The MR is positive and reaches a maximum value of 13\% at 2\,K and 80\,kOe. The MR is linear in field above about 30\,kOe for temperatures up to at least 100\,K. Approximately quadratic field dependence is seen at lower fields, typical of metals. However, the linear behavior above 30\,kOe persists up to 80\,kOe, and the expected saturation at high field is not seen in the temperature and field ranges studied here. Measurements to higher magnetic fields may be desirable to understand the origin of this linear, non-saturating magnetoresistance. Several sources of magnetoresistance behavior similar to this have been proposed, including high mobilities, linear band crossings, and fluctuations \cite{song2015, leahy2018, abrikosov1998, parish2005, hu2008}

Figure \ref{fig:normal}c and \ref{fig:normal}d show the results of magnetization measurements on PtTe crystals. The material is diamagentic from 2 to 300\,K. The plots also show data corrected for the core diamagnetism, giving the Pauli paramagentic behavior expected for this metallic compound. Isothermal magnetization curves measured at 2\,K are included in the Supplemental Material \cite{supp}. The measurements show a small amount of anisotropy, and the susceptibility is nearly temperature independent. A small Curie tail is seen below 25\,K. Fitting this upturn with a modified Curie Weiss law gives a Curie constant of $1\times10^{-4}$\,cm$^3$K/mol, corresponding to less than $3\times10^{-4}$ spin-1/2 centers per formula unit. The magnetization is nearly linear in field at 2\,K, as expected for a Pauli paramagnet. The small amount of curvature seen in isothermal magnetization curves (see Ref. \citenum{supp} may be attributed to the minute concentration of local moments indicated by the Curie tail. Taking the average of the two orientations, and neglecting the low temperature upturns, the Pauli susceptibility of PtTe can be estimated to be $\chi_0 = 7.0\times10^{-5}$\,cm$^3$/mol.

As expected based on the magnetization and resistivity behaviors, the heat capacity of PtTe shows no anomalies between 2 and 300\,K. This data, along with heat capacity data collected down to 0.63\,K in a 3He cryostat are plotted for low temperatures as $c_P/T$ vs $T^2$ in Fig. \ref{fig:normal}d. A linear fit from 0.63 to 2.3\,K give a Sommerfeld coefficient (intercept) of $\gamma$\,=\,3.0$\times10^{-3}$\,J/mol-FU/K$^2$ and a slope of 6.9$\times10^{-4}$J/mol-FU/K$^4$ corresponding to a Debye temperature of 140\,K.

With the normal state data presented here, the Wilson ratio can be calculated and provides a measure of correlated behavior, comparing the density of states inferred from the Pauli susceptibility with that inferred from the electronic heat capacity in the free electron model. The Wilson ratio is given by $R_W = \chi_0/\gamma$. For PtTe, in CGS units, $\gamma$\,=\,2.6$\times10^{4}$\,erg/mol-FU/K$^2$ and $\chi_0 = 7.0\times10^{-5}$\,emu/mol-FU/G, giving $R_W=2.7\times10^{-9}K^2/G^2$. This is about twice the free electron value of $\frac{3\mu_B^2}{\pi^2k_B^2} = 1.4\times10^{-9}K^2/G^2$. Such enhancement of the Wilson ratio can be a signature for correlations. However, the Sommerfeld coefficient calculated from the density of states at the Fermi level given by density functional theory (2.0$\times10^{-3}$\,J/mol-FU/K$^2$, see below), is close to the experimental value of 3.0$\times10^{-3}$\,J/mol-FU/K$^2$. This suggests no significant mass enhancement in PtTe.

\subsection{Superconducting properties}

\begin{figure}
\begin{center}
\includegraphics[width=3.5in]{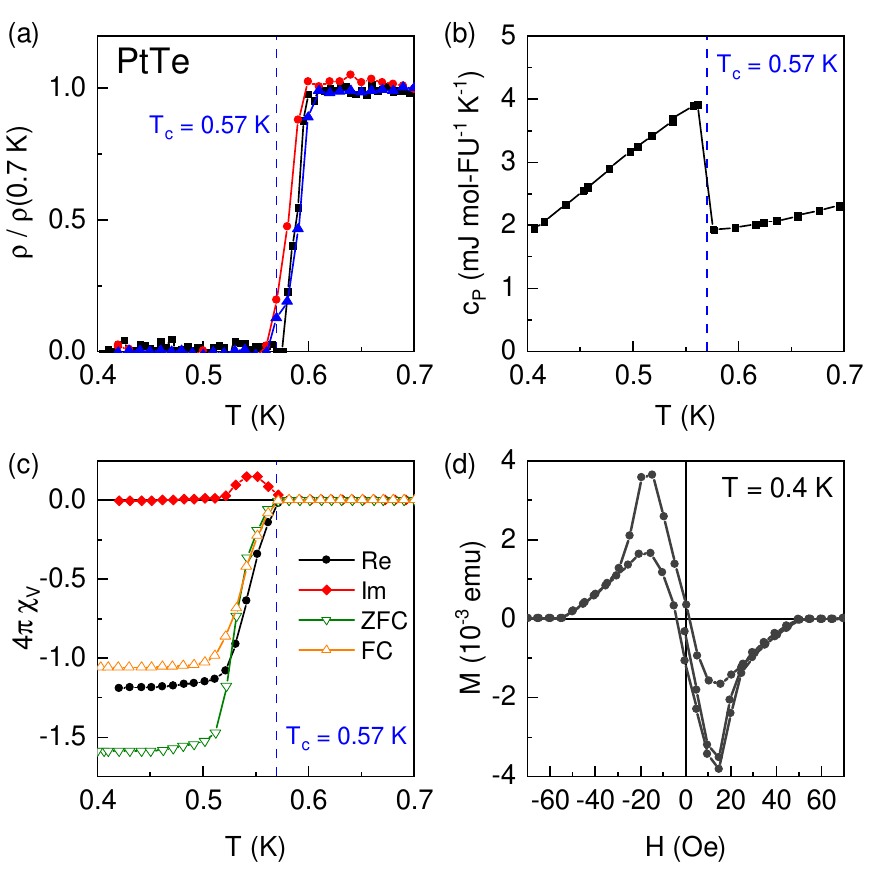}
\caption{\label{fig:SC-Tc}
Evidence for superconductivity below 0.57\,K in PtTe. (a) The electrical resistance of three PtTe crystals, normalized their values at 0.7\,K. (b) The specific heat capacity. (c) The real (Re) and imaginary (Im) parts of the ac magnetic susceptibility measured in zero applied dc field, and the dc magnetic susceptibility from zero-field-cooled (ZFC) and field-cooled (FC) measurements in a 5\,Oe applied field. (d) A magnetic hysteresis loop measured at 0.4\,K with the field in the ab plane.
}
\end{center}
\end{figure}

The behavior of PtTe at lower temperatures was investigated using a $^3$He cryostat, and revealed superconductivity with $T_c = 0.57$\,K. Unambiguous evidence for bulk superconductivity is presented in Fig. \ref{fig:SC-Tc}. A consisted transition temperature of 0.57\,K in zero field is determined by point at which the resistivity is zero, the onset of diamagnetism in magnetic susceptibility, and the midpoint of the heat capacity jump upon cooling through $T_c$.

The resistive transition is sharp (Fig. \ref{fig:SC-Tc}a) and the resistance reaches zero within the resolution of our measurement. The width of the transition is about 0.03\,K, or 5\% of the value of $T_c$. Data from three separate crystals measured in zero magnetic field are shown in the figure.

A clear heat capacity anomaly is observed at $T_c$, as shown in Fig. \ref{fig:SC-Tc}b. The heat capacity jump $\Delta c_P$ is measured to be 0.0020\,J\,mol-FU$^{-1}$K$^{-1}$. Using the Sommerfeld coefficient determined in Fig. \ref{fig:normal}, $\Delta c_P / \gamma T = 1.18$. This is close to the weak coupling BCS value of 1.43. The current data extends only to about 0.7T$_c$, and measurements at lower temperatures would be needed for further analysis of the entropy and gap structure.

The ac magnetic susceptibility data (Fig. \ref{fig:SC-Tc}c) show strong diamagnetism in the real part and a peak below $T_c$ in the imaginary part. The data in the figure were measured in zero applied dc field, with an ac drive frequency of 103\,Hz and an amplitude of 5\,Oe. The diamagnetic susceptibility reaches (and slightly exceeds) the expected value of 1/4$\pi$ in the superconducting state. The dc magnetic susceptibility in a 5\,Oe field from both field-cooled (FC) and zero-field-cooled (ZFC) measurements are also shown on the figure, and demonstrate strong diamagnetism below $T_c$. The divergence of FC and ZFC data in the supercondcuting state suggest PtTe to be a type-II superconductor. This is also supported by the magnetic hysteresis loop shown in Fig. \ref{fig:SC-Tc}d. The position of the minimum on increasing the field (or maximum on decreasing the field) and the approach to zero at higher fields define the lower critical field $H_{c1}$ and the upper critical field $H_{c2}$, respectively. Values of $H_{c1} = 15$\,Oe and $H_{c2} = 55$\,Oe are estimated from the data.

\begin{figure}
\begin{center}
\includegraphics[width=3.5in]{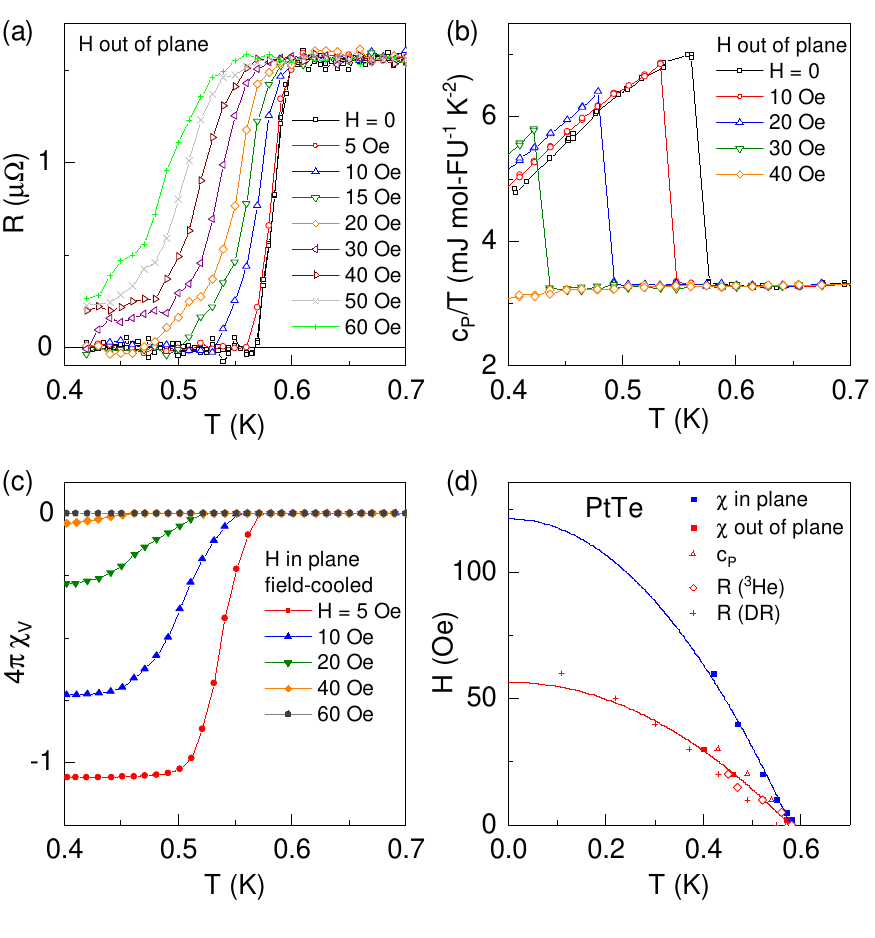}
\caption{\label{fig:SC-PD}
Effects of a magnetic field on the superconductivity in PtTe. (a) The electrical resistance in a field applied out of the plane (along the c axis). (b) The heat capacity plotted as $c_P/T$ measured with the field out of the plane. (c) The dc magnetic susceptibility with the field applied in the plane. (d) Superconducting phase diagram constructed from measurements on PtTe crystals, showing the temperature dependence of the upper critical field. The lines are fits using the phenomenological form $H_c(T) = H_c(0)[1-(\frac{T}{T_c})^2]$ and the data determined from the magnetic susceptibility.
}
\end{center}
\end{figure}

The superconducting phase diagram of PtTe was further studied by temperature dependent measurements of the electrical resistance, magnetization, and heat capacity in applied magnetic fields. The results are summarized in Fig. \ref{fig:SC-PD}. As noted above, $T_c$ is defined here by the point at which the resistance is zero, the onset of diamagnetism, and the midpoint of the heat capacity jump. Note that demagnetization effects in these plate like crystals may be significant when the field is directed out of the plane, but only when the diamagnetic susceptibility is large, that is below $T_c$. Thus, this should not affect the determination of $T_c$ values, since demagnetization effects are weak at the onset of diamagnetism. However, these effects would need to be considered in any analysis of the temperature or field dependence below $T_c$ for H out of the plane.

Electrical resistance measurements in fields up to 60\,Oe (current in the plane, magnetic field out of the plane) are shown in Fig. \ref{fig:SC-PD}a. The $T_c$ is suppressed below 0.4\,K by fields larger than about 30\,Oe. The transition is seen to broaden in applied fields, consistent with expectations for type-II superconductivity; however, this is complicated by the presence of a knee in the resistance curves apparent at fields above 10\,Oe. The data suggests some vestigial superconductivity, perhaps filamentary or surface, with a higher critical field coexists with the bulk superconductivity in PtTe in applied fields.

The heat capacity measured in out-of-plane magnetic fields up to 40\,Oe is shown in Fig. \ref{fig:SC-PD}b (plotted as $c_P/T$). A clear anomaly can be seen for fields of 30\,Oe and below. Higher fields suppress the transition below the lower limit of these measurements. There is also a weak anomaly in the heat capacity around 0.5\,K even in fields at which the bulk $T_c$ is suppressed below 0.4\,K. This can be seen in as a broad hump in the 40\,Oe data in Fig. \ref{fig:SC-PD}b. This is likely related to the vestigial superconductivity noted in the resistance data (see Supplemental Material \cite{supp} for additional data).

The effect of fields applied in the plane is shown in the magnetic susceptibility shown in Fig. \ref{fig:SC-PD}c. Higher critical fields are apparent in this orientation. Susceptibility measurements with the field out of the plane showed behavior consistent the heat capacity and resistance results in that same orientation.

Based on the temperature and field dependent measurements, the superconducting phase diagram can be constructed. Figure \ref{fig:SC-PD}d shows $H_{c2}$ vs $T$ for $H$ in the plane (susceptibility data) and out of the plane (susceptibility, resistance, and heat capacity data). The solid curves on the plot are fits using the phenomenological form $H_c(T) = H_c(0)[1-(\frac{T}{T_c})^2]$ and the data determined from the magnetic susceptibility. The Werthamer-Helfand-Hohenberg (WHH) theory gives an estimate of $H_{c2}(0) = -0.69T_c(dH_{c2}/dT)|_{T = T_c}$ for the critical field at zero temperature based on the slope near $T_c$ in low fields. Applying this WHH approximation gives estimates of $H_c(0)$ of 143\,Oe and 65\,Oe for fields in the plane and out of the plane, respectively. As a preliminary investigation of the behavior below 0.4\,K, resistance measurements were carried out in a dilution refrigerator down to 0.1\,K (see data in \citenum{supp}). Results from those measurements are included on the phase diagram (Fig. \ref{fig:SC-PD}d) and generally agree with the other results and the WHH critical field estimate.

\subsection{Electronic structure}

We examined the electronic properties of PtTe using density functional theory (DFT) calculations. For these calculations the primitive unit cell in the rhombohedral setting was used.
We note that our results are qualitatively consistent with the previous reports \cite{MaterialsProject} and \cite{TopologicalQuantumChemistry}.
We start from the structural optimization including the SOC. The optimized parameters are $a=7.2405$ {\AA} and $\alpha=32.1506^\circ$.
Our relaxed lattice constant $a$ is somewhat larger and our angle $\alpha$ is slightly smaller than both the literature value of $a=7.042$ {\AA} and $\alpha=32.685^\circ$ \cite{cenzual1990overlooked} and our experimental values of 6.990\,{\AA} and 32.921$^\circ$.
However, our optimized parameters are very close to the ones reported in \cite{MaterialsProject}.
A slight difference may be ascribed to the SOC included in this work.

\begin{figure}
\begin{center}
\includegraphics[width=3.5in]{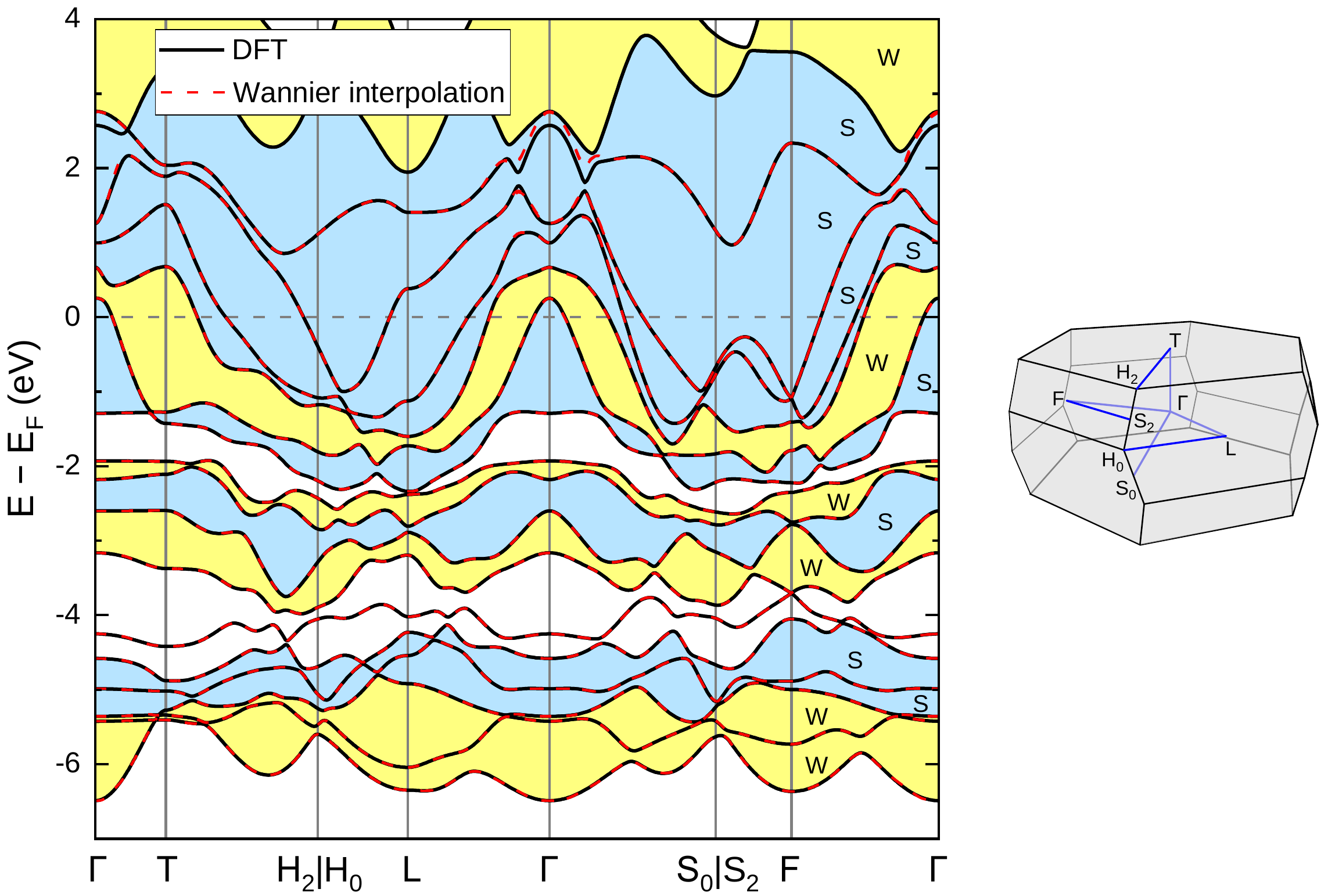}
\caption{DFT band structure calculated using the primitive rhombohedral cell. First Brillouin zone and high symmetry pointes are shown on the right. Due to the spatial inversion symmetry and the time reversal symmetry, each band is twofold degenerate.
Dispersion relations by Wannier interpolation are shown as red broken lines, showing an excellent agreement with DFT dispersions.
Shaded areas distinguish topological gaps from trivial gaps based on the parity eigenvalue analysis, where S (W) stand for strong (weak) TI, characterized by odd (even) number of surface Dirac bands.
}
\label{fig:band}
\end{center}
\end{figure}

The total and partial density of states from the DFT calculation show that Pt $d$ states and Te $p$ states mainly contribute to the electronic states around the Fermi level, as expected. The density of states are plotted shown in the Supporting Material \cite{supp}, and the electronic dispersions are shown in Fig. \ref{fig:band}.
Here, high-symmetry points are identified using SeekK-Path \cite{SeekKpath,Spglib}.
One notices that four hole-like bands cross the Fermi level. These bands form one small hole pocket centered at the $\Gamma$ point, two cylindrical FSs, and one large open FS (see Ref. \cite{supp}).
These results show a good agreement with the one reported in \cite{MaterialsProject} despite the difference in the inclusion of SOC.
The results presented in \cite{TopologicalQuantumChemistry} show a slight upward offset in the Fermi level by $\sim 1$ eV.
The calculated DOS at the Fermi level is close to 2\,eV$^{-1}$cell$^{-1}$. With two formula units per cell, this corresponds to $\gamma$\,=\,2.0$\times10^{-3}$\,J/mol-FU/K$^2$ in the free electron theory. As noted above, this is similar to the measured value of 3.0$\times10^{-3}$\,J/mol-FU/K$^2$.

Because of the spatial inversion symmetry and the time reversal symmetry, analyzing the topological nature of PtTe can be achieved by examining the parity eigenvalue of electronic bands at time-reversal invariant momentums (TRIMs) \cite{FuKane2007}. Strong and weak topological insulators are distinguished by $Z_2$ topological indexes, which are described and tabulated in the Supplemental Material \cite{supp}. Topological gaps are indicated in Fig. \ref{fig:band}, where band gaps showing strong and weak TI characters are indicated by S and W, respectively. Since the Fermi level crosses multiple topological gaps, a topological semimetallic state is realized in PtTe rather than a topological insulating state.

\begin{figure}
\begin{center}
\includegraphics[width=1\columnwidth, clip]{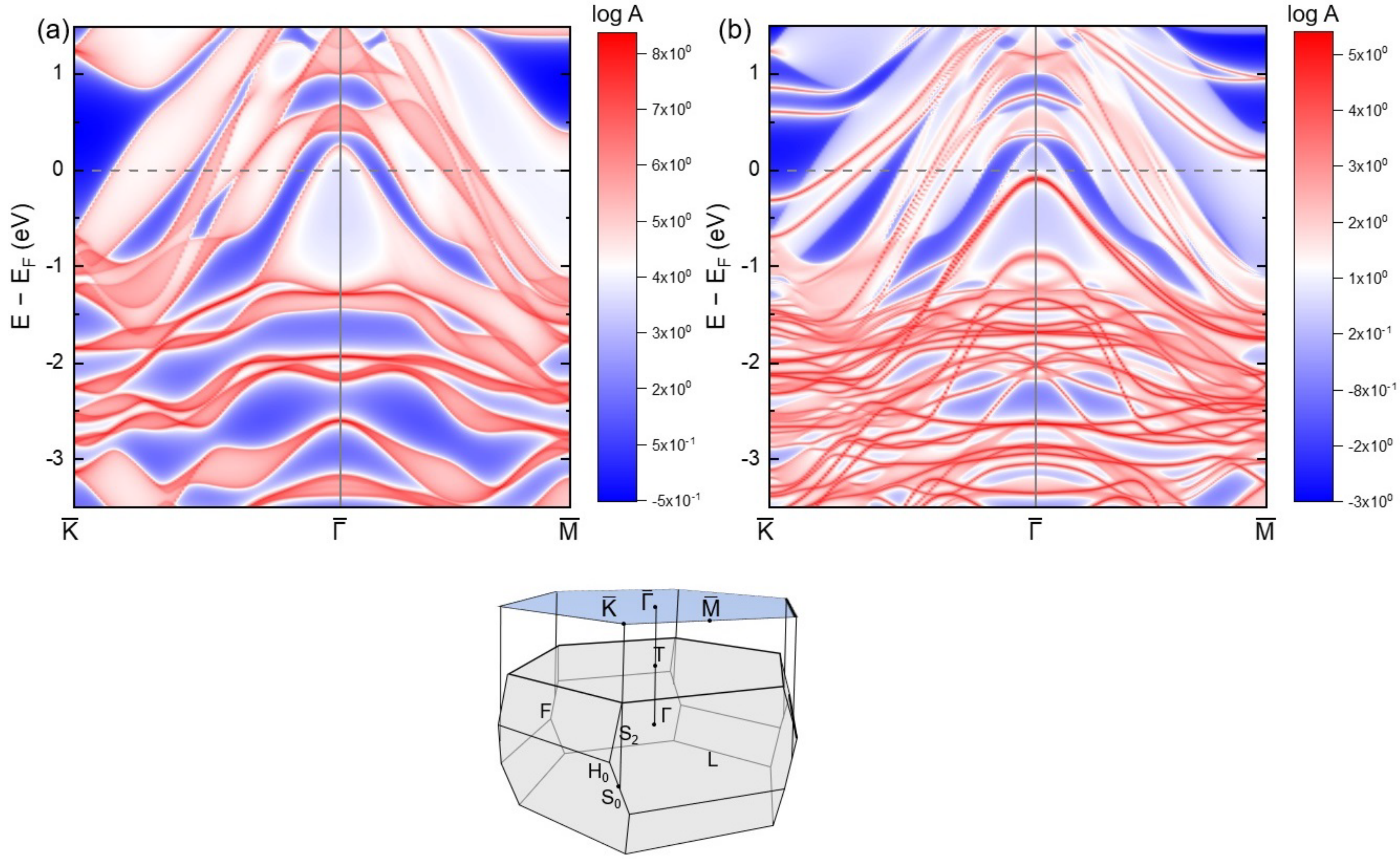}
\caption{Slab spectral functions. (a) bulk contribution and (b) surface contribution. Spectral intensity is given in the log scale.
The bottom figure shows the bulk and surface Brillouin zones.
}
\label{fig:Slab}
\end{center}
\end{figure}

\begin{figure}
\begin{center}
\includegraphics[width=1\columnwidth, clip]{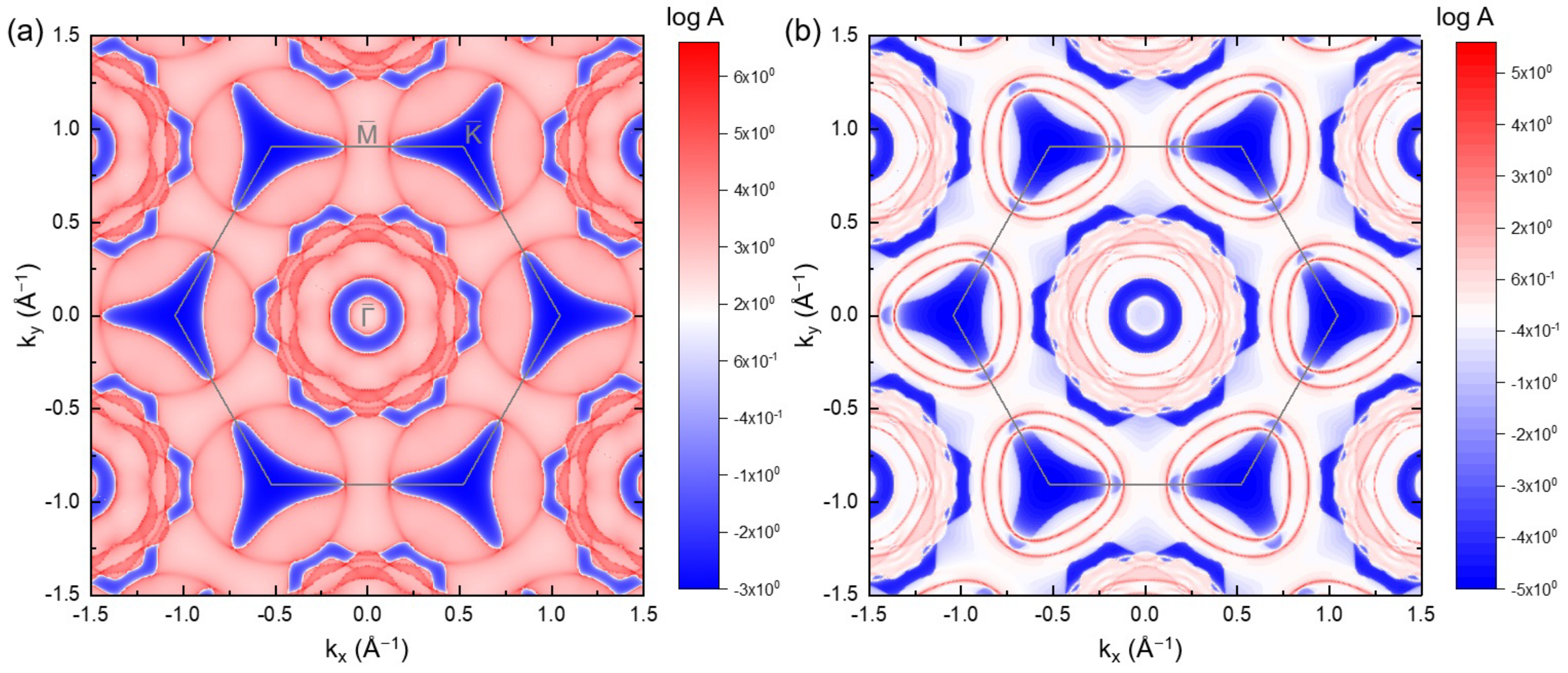}
\caption{Slab spectral functions on the Fermi surface. (a) bulk contribution and (b) surface contribution. Spectral intensity is given in the log scale.
The boundary of the surface Brillouin zone is indicated by gray lines.
}
\label{fig:SlabFermi}
\end{center}
\end{figure}

To further understand the topology of PtTe, we derived maximally localized Wannier functions projected onto Pt $d$ and Te $p$ states \cite{Wannier90}.
As shown in Fig. \ref{fig:band}, Wannier-interpolated band dispersion and the DFT dispersions show excellent agreement.
Using a Wannier tight-binding model, we compute the spectral functions using a finite thickness slab, where 20 unit-cells are stacked along the [111] direction \cite{WannierTools}.
Figs. \ref{fig:Slab} and \ref{fig:SlabFermi} show the slab spectral functions as functions of surface momentum and energy and those as functions of 2D surface momentum at the Fermi level, respectively.
On each figure, plot (a) shows the bulk contribution to the slab spectral function, and plot (b) shows the surface contribution.
Because of the highly dispersive bands, surface bands and bulk continuum overlap near the Fermi level. This makes the analysis of topological nature of the superconductivity challenging.
However, one notices a characteristic feature; dispersive surface bands come down above the Fermi level,
creating Dirac band crossing below the Fermi level, and come up above the Femi level.
This feature may be most visible in a surface band, which starts from the $\bar \Gamma$ point at $E-E_F \sim 1$ eV and create a Dirac crossing at the $\bar{\rm M}$ point at $E-E_F \sim -1.5$ eV.
Thus, this surface band crosses the Fermi level an odd number of times.
This is also seen in the surface spectral function at the Fermi level, Fig. \ref{fig:SlabFermi} (b),
where surface bands form double rings around the $\bar \Gamma$ and the $\bar {\rm K}$ points.
From this observation, the superconducting state in PtTe would have the trivial nature.

\begin{figure}
\begin{center}
\includegraphics[width=1\columnwidth, clip]{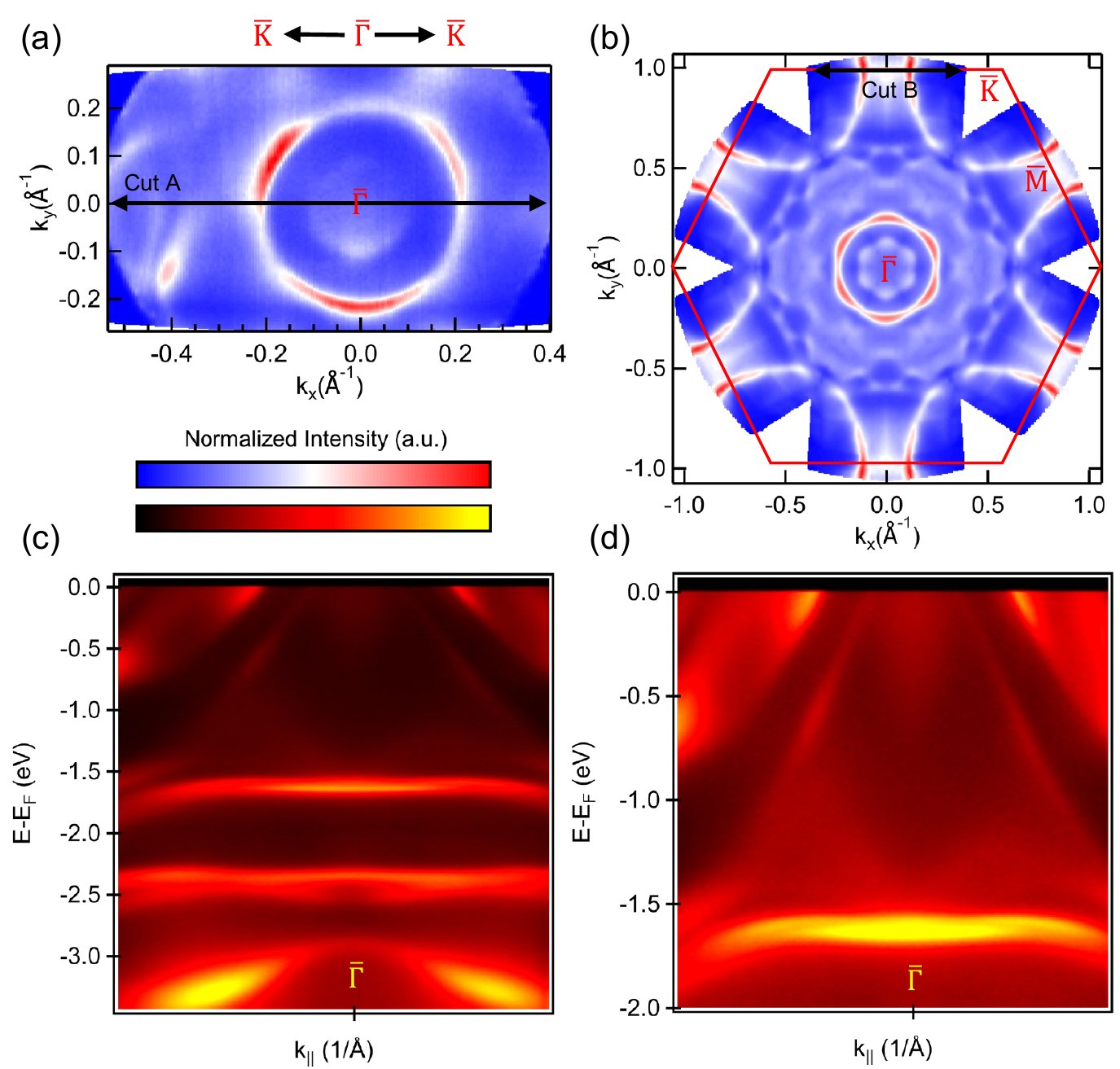}
\caption{ARPES Fermi surface and band structure at $\bar\Gamma$.  (a) Fermi surface centered at $\bar\Gamma$ using Scienta analyzer deflection mode.  Cut A indicates cut direction for band structure in (c) and (d).  (b) Symmetrized Fermi surface across first BZ by combining several several deflection mode maps.  BZ boundary and high symmetry points annotated on Map.  Cut B indicates cut direction for band structure in Fig.~\ref{fig:ARPES_M}. (c) Band structure along high symmetry $\bar{\rm K} - \bar\Gamma - \bar{\rm K}$ direction.  (d) Same data as in (c) but with adjusted intensity contrast to highlight weaker features near $E_F$. Different colormaps are chosen to help emphasize weaker features.}
\label{fig:ARPES_FS}
\end{center}
\end{figure}

We also investigated the electronic structure experimentally using ARPES. The Fermi surface and electronic structure along the high symmetry $\bar{\rm K} - \bar\Gamma - \bar{\rm K}$ direction for PtTe are shown in Fig.~\ref{fig:ARPES_FS}.  A small section of the Fermi surface centered about $\bar\Gamma$ obtained by steering the electrons across the analyzer slit within a $\pm 12^\circ$ window using the analyzer deflection mode are shown in Fig.~\ref{fig:ARPES_FS}a.  For ARPES, the maximum kinetic energy and parallel momentum of photoemitted electrons are set by the photon energy~\cite{Damascelli2003}.  With such a low the photon energy fixed at 11 eV, this translates into a measurement window with $k_{\vert\vert}\pm 0.3$ \AA$^{-1}$ centered at $\bar\Gamma$ along the analyzer deflection direction ($k_y$).  Even with a small window, we can clearly see two circular  pockets centered at $\bar\Gamma$ as well as many other spectral features at the Fermi level.  To reach higher $k_{\vert\vert}$ momenta for mapping out the entire BZ, we mechanically rotate the sample to a higher angle and again measure a small $k_{\vert\vert}$ window using the analyzer deflection mode.  These individual map sections are combined together and symmetrized based on the observed hexagonal symmetry to create a more complete Fermi surface map shown in Fig.~\ref{fig:ARPES_FS}b. The measured Fermi surface is in remarkable agreement with the calculated Fermi surface and confirms the metallic electronic structure with numerous bands crossing the Fermi level.

When looking at the electronic structure along the $\bar{\rm K} - \bar\Gamma - \bar{\rm K}$ high symmetry direction as shown in Fig.~\ref{fig:ARPES_FS}c, we see hole pockets centered at $\bar\Gamma$, consistent with both theory and Hall effect data.  The overall agreement with theoretical calculations including SOC is quite remarkable with minimal offset and band renormalization observed. This is also consistent with the lack of mass enhancement noted above. Upon closer inspection there is weak intensity just below the Fermi level that is not accounted for in the bulk band calculations.  However, the calculated spectral weight for the surface shows a band with a dispersion maximum just below the Fermi level and the observed intensity could be associated with this surface band.  Due to the limitations of the experimental base temperature of $\sim 7$ K compared with the superconducting $T_c = 0.57$ K, only the normal state electronic structure is measured.  As noted above, the enhanced Wilson ratio could be taken as a signature of correlations in PtTe. In ARPES data, strong correlations are often revealed by renormalized bands with interaction-induced mass enhancements and a reduced spectral weight plus increases in dispersion linewithdts below the Fermi level due to finite lifetime~\cite{Damascelli2003,Mackenzie2003,Mravlje2011}.  However, dispersing bands remain sharp away from $E_F$ and as noted above, minimal band renormalization is observed when compared to theoretical calculations. Such observations show that PtTe is likely far away from a strongly correlated regime.

\begin{figure}
\begin{center}
\includegraphics[width=1\columnwidth, clip]{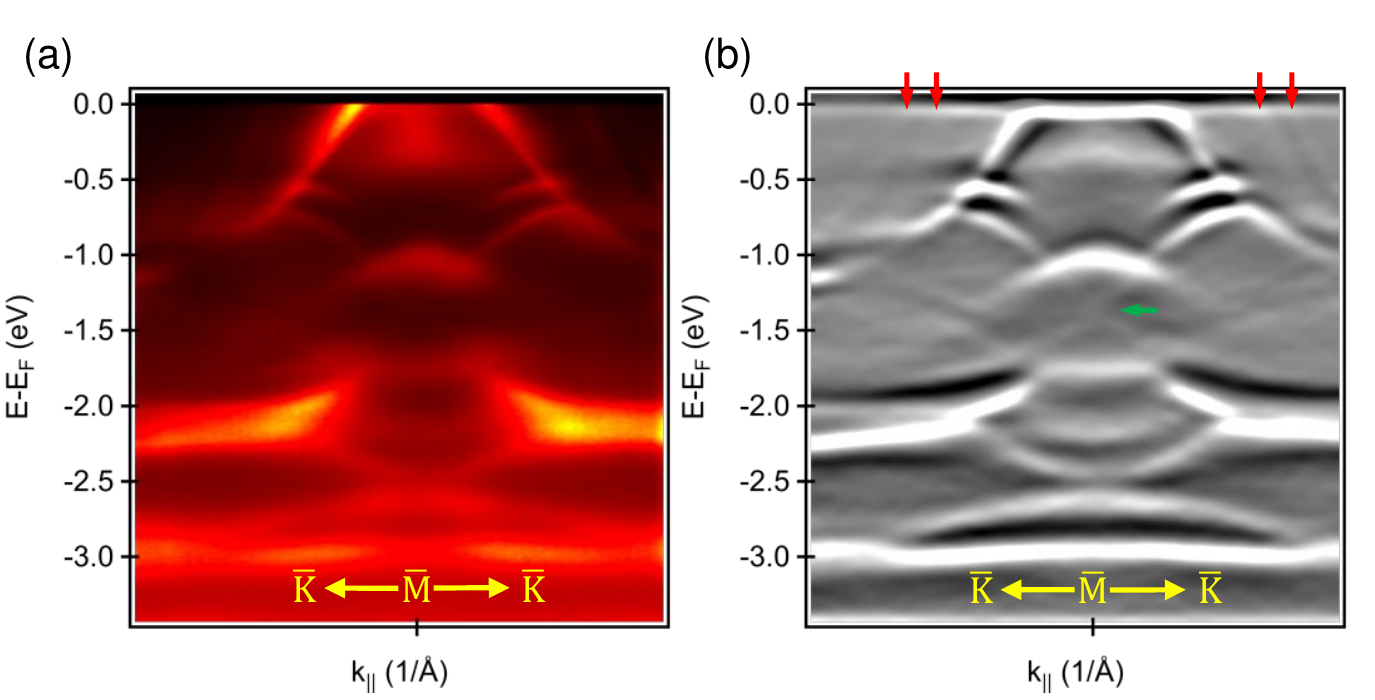}
\caption{ARPES band structure at $\bar {\rm M}$ point.  (a) Electronic structure along the $\bar {\rm K} - \bar {\rm M} - \bar {\rm K}$ direction.  Location in BZ is indicated by Cut B in Fig.~\ref{fig:ARPES_FS}b.  (b) Curvature analysis of data in (a).  Red arrows highlight weak intensity bands crossing the Fermi level.  Green arrow highlights surface bands crossing.}
\label{fig:ARPES_M}
\end{center}
\end{figure}

From our theoretical investigation into the topological nature of PtTe we find the system is a topological semimetal with numerous surface bands.  To investigate the topological states, we focus on the $\bar{\rm M}$-point where several topological bands should be observed.  The electronic structure along the $\bar{\rm K} - \bar{\rm M} - \bar{\rm K}$ direction are shown in Fig.~\ref{fig:ARPES_M}.  When comparing the theoretical spectral weight plots for the bulk and surface contributions to the Fermi surface, when progressing along the Brillouin zone boundary from  $\bar{\rm M}$ to $\bar{\rm K}$, we should expect to see a single Fermi level crossing from bulk bands as well as two band crossings from the surface bands.  In the ARPES data shown in Fig.~\ref{fig:ARPES_M}a a clear hole-like pocket centered at $\bar{\rm M}$ is observed with weaker spectral features observed further from the  $\bar{\rm M}$-point.  To better visualize the weaker spectral features, a plot of the data using a curvature analysis is shown in Fig.~\ref{fig:ARPES_M}b~\cite{Zhang2011}.  From the curvature plot, two additional bands are observed crossing the Fermi level between the  $\bar{\rm M}$ and  $\bar{\rm K}$-points as highlighted with red arrows in Fig.~\ref{fig:ARPES_M}b.  From the theoretical spectral weight plots, one of the surface band crossings occurs at the same location along the high symmetry $\bar{\rm M} - \bar{\rm K}$ line as the bulk band crossings, however, we can resolve three crossings in the experimental data.  In addition the the band crossings at the Fermi level, the curvature analysis shows two bands crossing at the $\bar{\rm M}$-point near $E-E_F \sim -1.4$ eV that is highlighted by the green arrow, which is again consistent with theoretical calculations.  These combination of theory and ARPES confirms the topological nature of the PtTe system.

The intensity of these additional bands, attributed to surface bands, is very weak and while observed in the raw data, are more clear in the curvature plot.  The spectral feature just below the $\bar\Gamma$ point, which may be attributed to a surface band is also very weak.  The ARPES data from many topological materials as Bi$_2$Te$_3$ ~\cite{Chen2009, Chen2010}, Bi$_2$Se$_3$~\cite{Analytis2010}, and Bi$_{1-x}$Sb$_x$ ~\cite{Hsieh2009} clearly resolve the topological surface states with intensity comparable with bulk bands.  However, it has been shown that the topological surface band intensity varies dramatically with photon energy and polarization~\cite{Bao2012, Zhang2018}. Further synchrotron investigations with different photon energies and polarizations may help highlight the surface bands in this material.

\section{Summary and conclusions}

PtTe is a van der Waals layered compound that is relatively easily grown as plate-like single crystals using the flux technique. The crystals are easily cleaved and exfoliated. It is superconducting, with a critical temperature of 0.57\,K and relatively low critical fields. The behavior in a magnetic field is consistent with type-II superconductivity. Electronic structure calculations and angle resolved photoemission spectroscopy reveal topological surface states in PtTe; however, the superconducting state is expected to be topologically trivial.

In considering future work based on this study, we note that chemical tuning through alloying or doping often provides a route to enhanced or new functionality. Often this starts with exploring alloys that combine components of isostructural compounds. In the case of PtTe, the only isostructural phases are halides, and while these are structurally analogous the chemistry is not particularly similar. Still it may be interesting to study the effects of partially replacing Pt with other transition metals, particularly with Pd or Ni, to see how the superconductivity might be affected, or with more strongly magnetic transition metals in an effort to induce magnetic order.

\section*{Acknowledgements}
This material is based upon work supported by the U.S. Department of Energy, Office of Science, National Quantum Information Science Research Centers, Quantum Science Center.


%


\clearpage
\newpage

\setcounter{equation}{0}
\setcounter{figure}{0}
\setcounter{table}{0}
\setcounter{page}{1}
\makeatletter
\renewcommand{\theequation}{S\arabic{equation}}
\renewcommand{\thefigure}{S\arabic{figure}}
\renewcommand{\thetable}{S\arabic{table}}
\renewcommand{\bibnumfmt}[1]{[S#1]}
\renewcommand{\citenumfont}[1]{S#1}

\section{Supplementary Materials}

\begin{figure}
\begin{center}
\includegraphics[width=3.5in]{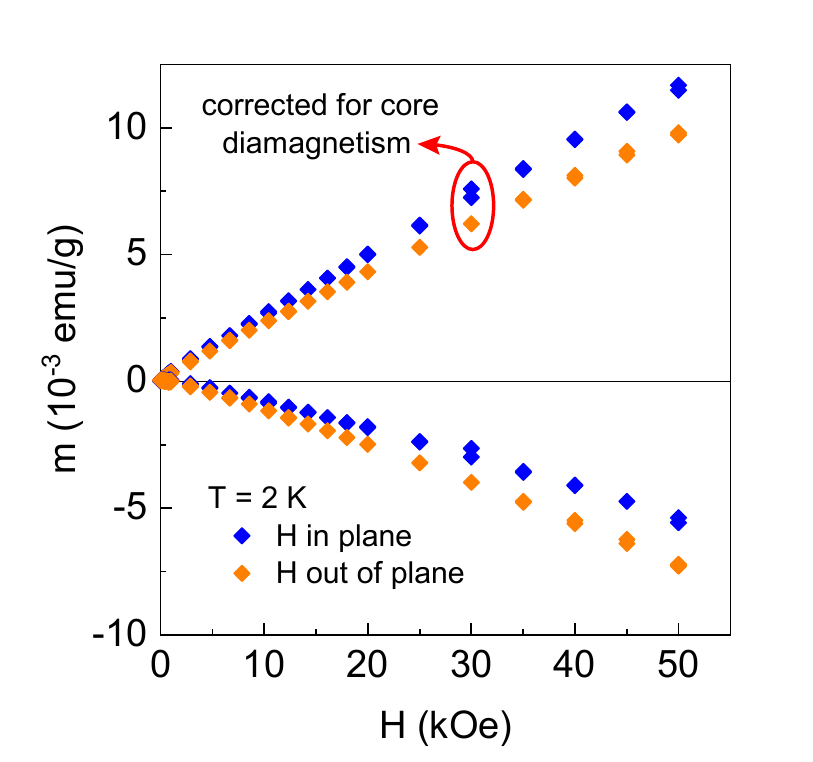}
\caption{\label{fig:MH2K}
Normal state M vs H data for PtTe, measured at 2\,K. The total moment and the moment corrected for core diamagnetism are both shown for both orientations.
}
\end{center}
\end{figure}

\begin{figure}
\begin{center}
\includegraphics[width=3.5in]{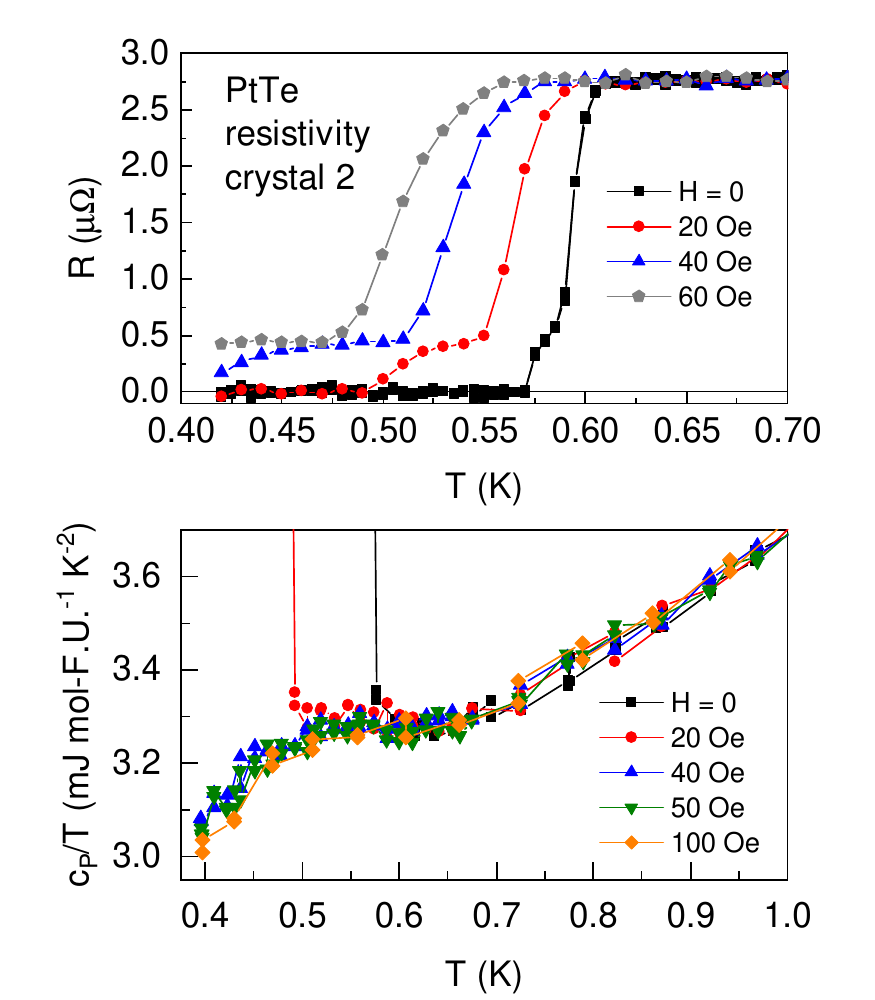}
\caption{\label{fig:vestigial-SC}
Resistance and heat capacity data from PtTe showing evidence for vestigial superconductivity with a higher critical field than the bulk.
}
\end{center}
\end{figure}

\begin{figure}
\begin{center}
\includegraphics[width=3.5in]{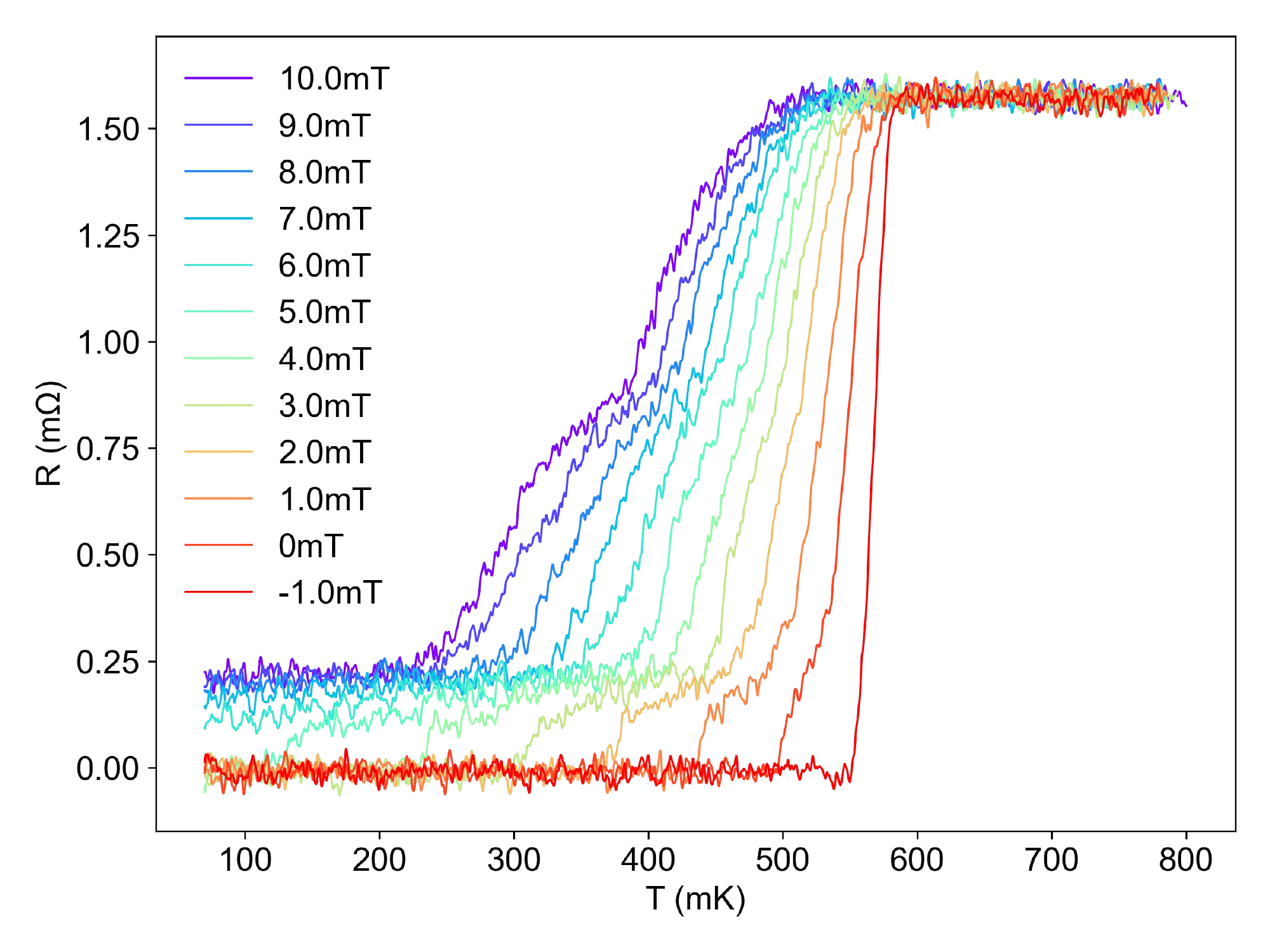}
\caption{\label{fig:DR-res}
Data from dilution refrigerator measurements of resistance vs temperature from a PtTe crystal in magnetic fields applied out of the plane. The maximum $T_c$ is observed at a nominal applied field value of -1 mT (10 Oe) due to residual trapped field in the magnet.
}
\end{center}
\end{figure}

Crystallographic details are collected in Tables \ref{tab:scxrd} and \ref{tab:adp}. Isothermal magnetization curves measured at 2\,K are shown in Fig. \ref{fig:MH2K}. Evidence for vestigial superconductivity in applied fields is shown in Fig. \ref{fig:vestigial-SC}. Low temperature resistance data are shown in Fig. \ref{fig:DR-res}.

\begin{figure}
\begin{center}
\includegraphics[width=3.5in]{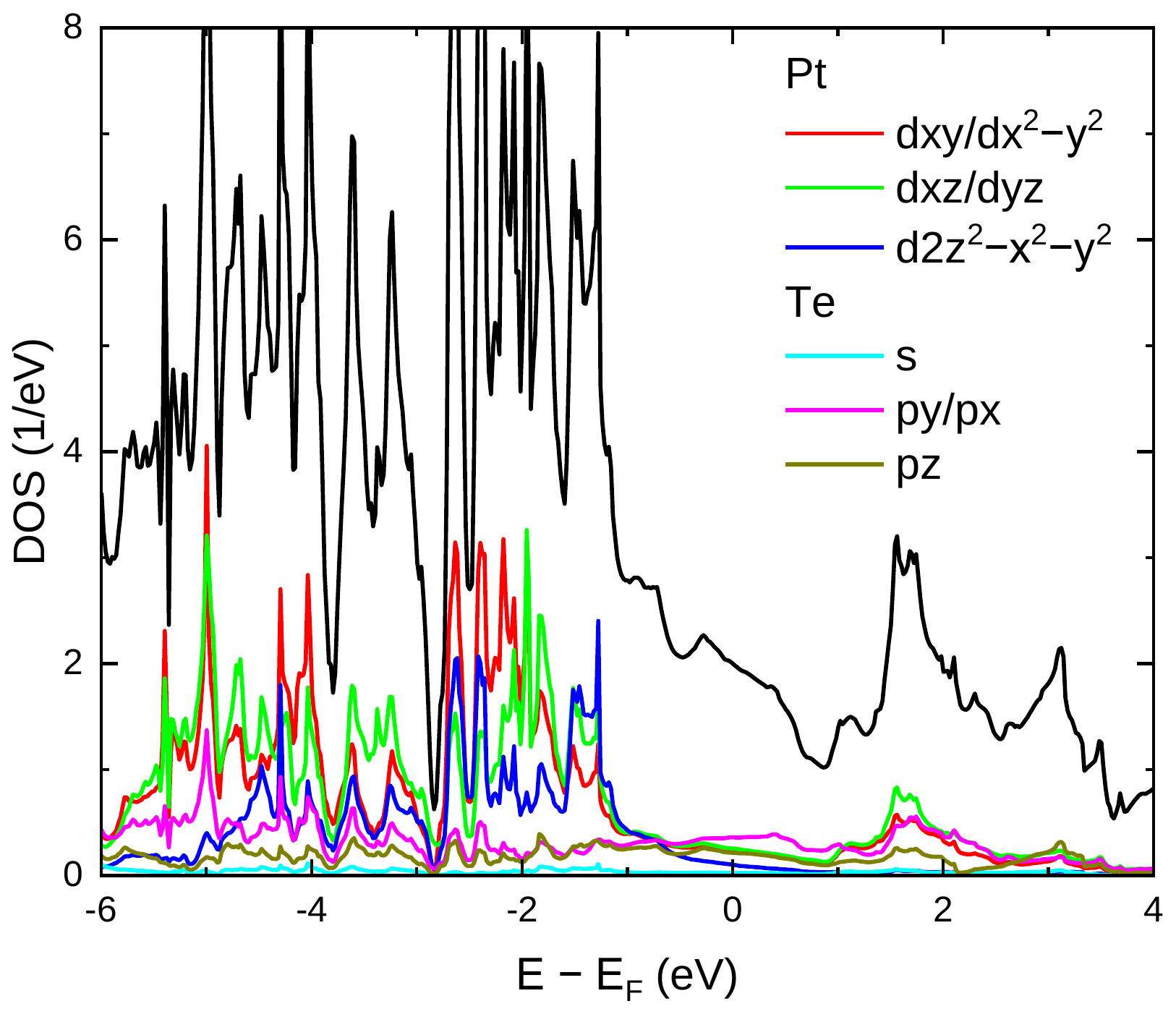}
\caption{Total density of states (DOS, black line) and partial DOS projected onto Pt d and Te s and p states.
}
\label{fig:DOS}
\end{center}
\end{figure}

\begin{figure}
\begin{center}
\includegraphics[width=3.5in, clip]{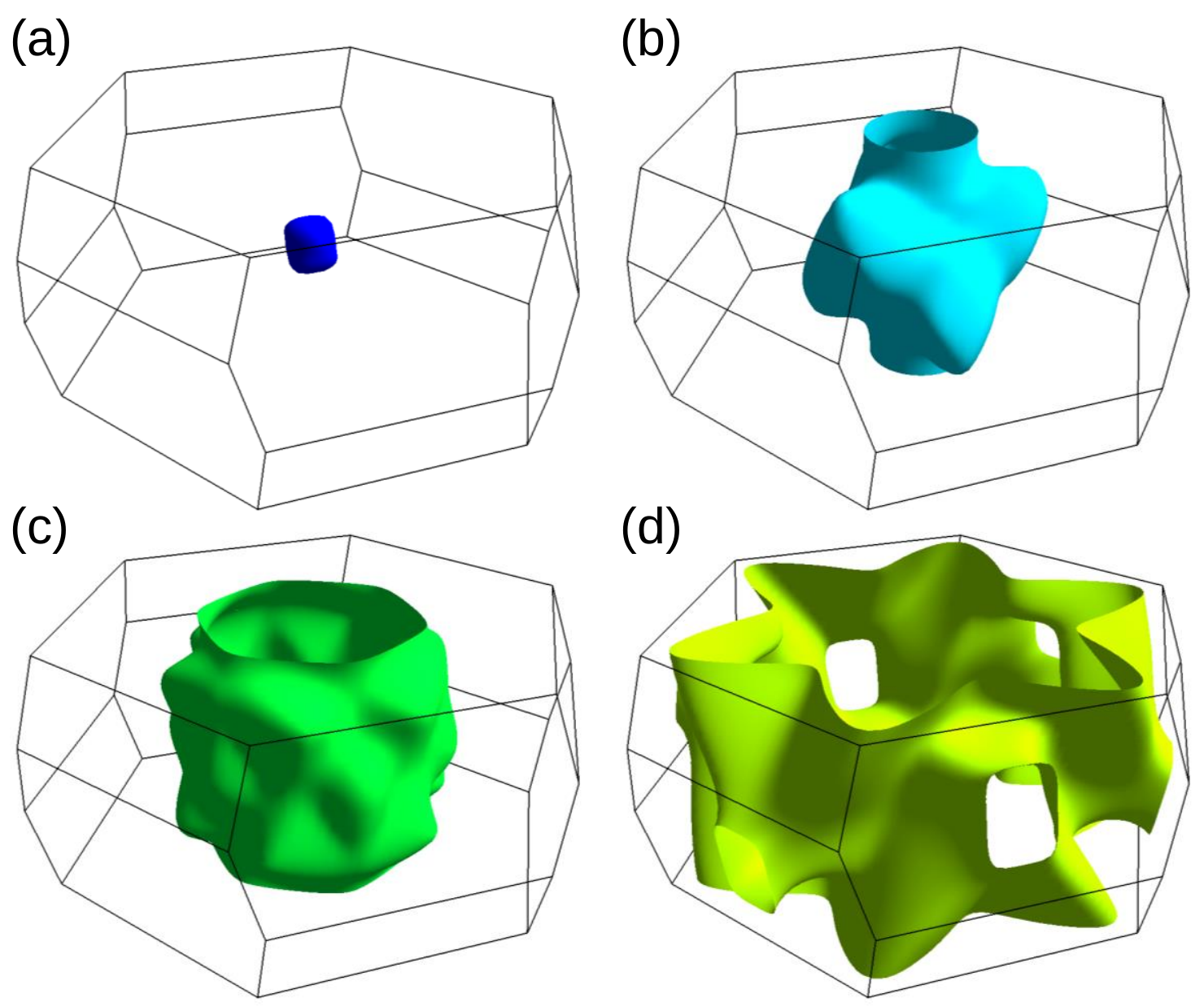}
\caption{Four Fermi surfaces. There appear one hole pocked centered at the $\Gamma$ point (a), two hole like cylindrical FSs (b,c), and one large open FS (d).
The FSs are computed using Wannier90 package \cite{Wannier90} and visualized using Fermisurfer package \cite{Fermisurfer}.
}
\label{fig:Fermi}
\end{center}
\end{figure}

The calculated density of states, total and projected, are shown in Fig. \ref{fig:DOS}, with the Fermi surface shown in Fig. \ref{fig:Fermi}.

Parity eigenvalues determined using the irvsp package \cite{Irvsp} are summarized in Table \ref{tab:parity}.
Here, the $Z_2$ topological indexes $(\nu_0; \nu_1, \nu_2, \nu_3 )$ are also presented.
For this, first TRIMs are represented as ${\boldsymbol \Gamma}_{n_1 n_1 n_3}=1/2 (n_1 {\boldsymbol b}_1+n_2 {\boldsymbol b}_2+n_3 {\boldsymbol b}_3 )$,
where $n_{1,2,3}=0,1$ and ${\boldsymbol b}_{1,2,3}$ are reciprocal lattice vectors.
Using these $n_{1,2,3}$ and corresponding parity eigenvalue $\delta_{n_1 n_2 n_3}$ at a given TRIM,
$\nu_i$ is given by $(-1)^{\nu_0}=\prod_{n_j=0,1} \delta_{n_1 n_2 n_3}$ and
$(-1)^{\nu_{i=1,2,3}}=\prod_{n_{j \ne i}=0,1,n_i=1} \delta_{n_1 n_1 n_3}$.
Finally, all $\nu_i$ are summed up below the focused band and divided by 2 and take its reminder.

\begin{table}
\caption{\label{tab:scxrd} Results from refinement of single crystal x-ray diffraction data from PtTe.}
\setlength{\tabcolsep}{2mm}
\begin{tabular}{lc}					
\hline
Empirical formula	&	Pt Te	\\
Formula weight	&	322.69	\\
Temperature	&	200(2) K	\\
Wavelength	&	0.71073 {\AA}	\\
Crystal system, space group	&	Trigonal,  R -3 m :H	\\
Unit cell dimensions	&	a = 3.9612(4) {\AA}   $\alpha$ = 90 deg.	\\
	&	b = 3.9612(4) {\AA}    $\beta$ = 90 deg.	\\
	&	c = 19.815(2) {\AA}   $\gamma$ = 120 deg.	\\
Volume	&	269.27(6) {\AA}$^3$	\\
Z, Calculated density	&	6,  11.940 Mg/m$^3$	\\
Absorption coefficient	&	93.499 mm$^{-1}$	\\
F(000)	&	780	\\
Crystal size	&	0.290 x 0.080 x 0.010 mm	\\
Theta range for data collection	&	3.084 to 38.980 deg.	\\
Limiting indices	&	-6$\leq$h$\leq$6, -6$\leq$k$\leq$6, -34$\leq$l$\leq$16	\\
Reflections collected / unique	&	1318 / 223 [R(int) = 0.0661]	\\
Completeness to theta = 25.242	&	0.988	\\
Absorption correction	&	Semi-empirical from equivalents	\\
Max. and min. transmission	&	0.5691 and 0.1144	\\
Refinement method	&	Full-matrix least-squares on F$^2$	\\
Data / restraints / parameters	&	223 / 0 / 9	\\
Goodness-of-fit on F$^2$	&	1.141	\\
Final R indices [I$>$2$\sigma$(I)]	&	R1 = 0.0705, wR2 = 0.1662	\\
R indices (all data)	&	R1 = 0.0709, wR2 = 0.1663	\\
Extinction coefficient	&	0.013(2)	\\
Largest diff. peak and hole	&	14.793 and -10.477 e {\AA}$^{-3}$	\\
\hline				
\end{tabular}	
\end{table}
\begin{table}
\caption{\label{tab:adp} Anisotropic displacement parameters ({\AA}$^2 \times 10^3$) from refinement of single crystal x-ray diffraction data from PtTe. The anisotropic displacement factor exponent takes the form:
   $-2\pi^2 [ h^2 a^{*2} U_{11} + ... + 2 h k a^* b^* U_{12} ]$}
\setlength{\tabcolsep}{3mm}
\begin{tabular}{lcccccc}					
\hline
	&	U$_{11}$	&	U$_{22}$	&	U$_{33}$	&	U$_{23}$	&	U$_{13}$	&	U$_{12}$	\\
\hline													
Pt	&	8(1)	&	8(1)	&	5(1)	&	0	&	0	&	4(1)	\\
Te	&	9(1)	&	9(1)	&	5(1)	&	0	&	0	&	4(1)	\\
\hline													

\hline				
\end{tabular}	
\end{table}

\begin{table}
\caption{Parity eigenvalues of electronic bands at TRIMs and resulting $Z_2$ indexes.
The first column shows the label of bands in the increasing order of energy from the  lowest band.
Only odd labels are shown because of the twofold degeneracy due to the time-reversal and spatial inversion symmetry.
From the second to the seventh columns, energy in the unit of eV and the parity eigenvalues of given TRIM are shown.
The last column shows the band-dependent $Z_2$ index. $\times 3$ indicates the number of equivalent TRIMs. }
\begin{center}
\begin{tabular}{ |c|cc|cc|cc|cc|c |}
 \hline
& \multicolumn{2}{c|}{$\Gamma$} & \multicolumn{2}{c|}{L $\times 3$} & \multicolumn{2}{c|}{F $\times 3$} & \multicolumn{2}{c|}{T } & $(\nu_0;\nu_1,\nu_2\nu_3)$ \\
\hline
1 & $-12.97$ & $+$ & $-11.69$ & $+$ & $-11.48$ & $-$ & $-12.43$ & $+$ & $(1;0,0,0)$ \\
3 & $-11.66$ & $-$ & $-11.24$ & $-$ & $-11.47$ & $+$ & $-12.41$ & $-$ & $(0;0,0,0)$ \\
5 & $-6.49$ & $+$ & $-6.35$ & $-$ & $-6.37$ & $-$ & $-5.41$ & $+$ & $(0;1,1,1)$ \\
7 & $-5.43$ & $+$ & $-6.04$ & $+$ & $-5.73$ & $+$ & $-5.34$ & $+$ & $(0;1,1,1)$ \\
9 & $-5.36$ & $+$ & $-4.92$ & $-$ & $-5.00$ & $-$ & $-5.27$ & $-$ & $(1;1,1,1)$ \\
11 & $-4.99$ & $-$ & $-4.54$ & $+$ & $-4.88$ & $+$ & $-5.02$ & $-$ & $(1;0,0,0)$ \\
13 & $-4.58$ & $-$ & $-4.23$ & $+$ & $-4.05$ & $+$ & $-4.88$ & $+$ & $(0;0,0,0)$ \\
15 & $-4.25$ & $-$ & $-4.02$ & $-$ & $-3.71$ & $-$ & $-4.42$ & $-$ & $(0;0,0,0)$ \\
17 & $-3.17$ & $-$ & $-3.19$ & $+$ & $-3.70$ & $+$ & $-3.37$ & $-$ & $(0;1,1,1)$ \\
19 & $-2.60$ & $-$ & $-2.89$ & $+$ & $-2.78$ & $-$ & $-2.59$ & $-$ & $(1;1,1,1)$ \\
21 & $-2.18$ & $+$ & $-2.81$ & $-$ & $-2.75$ & $+$ & $-2.11$ & $+$ & $(0;1,1,1)$ \\
23 & $-1.93$ & $+$ & $-2.38$ & $-$ & $-2.35$ & $-$ & $-1.93$ & $+$ & $(0;0,0,0)$ \\
25 & $-1.29$ & $+$ & $-2.34$ & $+$ & $-2.22$ & $+$ & $-1.43$ & $-$ & $(1;1,1,1)$ \\
27 & $0.26$ & $+$ & $-1.72$ & $+$ & $-1.79$ & $-$ & $-1.27$ & $+$ & $(0;1,1,1)$ \\
29 & $0.67$ & $-$ & $-1.60$ & $-$ & $-1.41$ & $-$ & $0.68$ & $+$ & $(1;0,0,0)$ \\
31 & $1.00$ & $+$ & $-1.12$ & $-$ & $-1.49$ & $-$ & $1.51$ & $+$ & $(1;1,1,1)$ \\
33 & $1.26$ & $+$ & $0.38$ & $-$ & $-1.08$ & $+$ & $1.89$ & $-$ & $(1;1,1,1)$ \\
35 & $2.57$ & $-$ & $1.41$ & $+$ & $2.34$ & $+$ & $2.04$ & $-$ & $(1;0,0,0)$ \\
37 & $2.76$ & $-$ & $1.95$ & $-$ & $3.36$ & $-$ & $3.32$ & $+$ & $(0;1,1,1)$ \\
39 & $4.53$ & $-$ & $3.61$ & $+$ & $4.63$ & $+$ & $4.38$ & $-$ & $(1;0,0,0)$ \\
\hline
\end{tabular}
\label{tab:parity}
\end{center}
\end{table}

\end{document}